\title{Properties of the weighted log-rank test in the design of confirmatory studies with delayed effects}
\author{Jos\'e L. Jim\'enez$^1$, Viktoriya Stalbovskaya$^2$ and Byron Jones$^3$\\
$^1$Politecnico di Torino, Turin, Italy\\
$^2$Merus, Utrecht, The Netherlands\\
$^3$Novartis Pharma A.G., Basel, Switzerland}

\date{}

\documentclass[12pt]{article}

\usepackage{amsmath,amssymb}
\usepackage{bbm}
\usepackage[a4paper, total={7in, 10in}]{geometry}
\usepackage{graphicx}
\usepackage{makecell}
\usepackage{bigints}
\usepackage{multirow}
 \usepackage{relsize}
\usepackage{algorithm, algpseudocode}
\usepackage{bbm}
\usepackage{multirow}
\usepackage{amsmath,amsfonts,amssymb,amsthm,epsfig,epstopdf,titling,url,array}
\usepackage{flexisym}
\usepackage{algorithm}
\usepackage{algpseudocode}
\usepackage{mathtools}
\usepackage{listings}
\usepackage[svgnames]{xcolor}

\usepackage{color}
\newcommand{\com}[1]{{\color{black} #1}}

\usepackage{ctable}

\newcommand\myeq{\stackrel{\tiny\mathclap{\mbox{appx.}}}{\approx}}

\begin{document}
\maketitle

\begin{abstract}

Proportional hazards are a common assumption when designing confirmatory clinical trials in oncology. This assumption not only affects the analysis part but also the sample size calculation. The presence of delayed effects causes a change in the hazard ratio while the trial is ongoing since at the beginning we do not observe any difference between  treatment arms and after some unknown time point, the differences between treatment arms will start to appear. Hence, the proportional hazards assumption no longer holds and both sample size calculation and analysis methods to be used should be reconsidered. The weighted log-rank test allows a weighting for early, middle and late differences through the Fleming and Harrington class of weights, and is proven to be more efficient when the proportional hazards assumption does not hold. The Fleming and Harrington class of weights, along with the estimated delay, can be incorporated into the sample size calculation in order to maintain the desired power once the treatment arm differences start to appear. In this article, we explore the impact of delayed effects in group sequential and adaptive group sequential designs, and make an empirical evaluation in terms of power and type-I error rate of the of the weighted log-rank test in a simulated scenario with fixed values of the Fleming and Harrington class of weights. We also give some practical recommendations regarding which methodology should be used in the presence of delayed effects depending on certain characteristics of the trial.

\vspace{0.5cm}

Keywords: adaptive designs; confirmatory trials; delayed effects; immuno-oncology agents; weighted log-rank.

\end{abstract}

\section{Introduction}

In drug development, randomized controlled trials remain the gold standard to confirm efficacy and safety of novel drug candidates. Often phase III trials embed formal interim analyses to allow studies to be stopped earlier for futility if the novel drug is not efficacious or for efficacy if the treatment effect is overwhelmingly positive.

Immuno-oncology (IO) is a rapidly evolving area in the development of anti-cancer drugs. IO agents can have effect on both the human immune system and the tumor microenvironment. By doing so, the tumors may be eradicated from the host or disease progression may be delayed. The effect of an IO agent is not typically directed to the tumor itself; it instead boosts or releases the brake from the patient’s immune system, and this positive effect may not be observed immediately. The lag between the activation of immune cells, their proliferation and impact on the tumor is described in the literature as a delayed treatment effect. Some patients may not derive clinical benefit before their disease progresses while others may derive sustained response or control of their disease. The primary endpoints often used for confirmatory phase III studies in oncology are time to event: progression free survival (PFS) and overall survival (OS). PFS is defined as time from randomization until disease progression or death and OS is defined as time from randomization until death from any cause. The delayed treatment effect may translate to inferior or equal PFS or OS compared to control treatment in the first months of therapy and superior survival thereafter leading to non-proportionality of hazards in the experimental and control arms of study. Therefore, the original design based on a proportional hazards assumption will lead to an underpowered study and hence both the sample size calculation and the analysis methods to be used should be reconsidered \cite{chen2013statistical}.

A weighted version of the log-rank test that incorporates the Fleming and Harrington class of weights \cite{fleming1981class}, allows tuning the two parameters $(\rho, \gamma)$ depending on if we expect early, middle or late delays, is proposed in the literature to increase the power at the end of the trial. However, tuning these parameters is not straightforward, since a misspecification may cause an even larger power drop with respect to the log-rank test.

The Fleming and Harrington class of weights, along with the estimated delay, can be incorporated into the sample size calculation in order to maintain the desired power once the treatment arm differences start to appear (see \cite{hasegawa2014sample}).

In this article we make an empirical evaluation of the impact of having a delayed effect on power and type I error rate in the design of a confirmatory phase III study with an IO agent used in combination with a standard of care, assuming a range for delay time. We assess the performance of the weighted log-rank test as an alternative to the log-rank test given it allows weighting of late differences and the potential gain power under non-proportional hazards. The evaluation is made for both group sequential and adaptive group sequential designs with fixed values of the Fleming and Harrington class of weights. We also give some practical recommendations regarding the methodology to be used in the presence of delayed effects depending on certain characteristics of the trial.

The manuscript is organized as follows. In section 2, we describe the weighted log-rank test and derive derive the sample size calculation formula needed to incorporate the estimated delay and the Fleming and Harrington class of weights, and we introduce the combination test statistic that will be necessary when doing sample size re-assessment. In section 3 we briefly describe group sequential and adaptive group sequential designs, emphasizing two popular methods used to do sample size re-assessment. In section 4, we describe the simulated example.

\section{Methods}
\label{sectionMethods}

In this section we describe the statistical methodology we review in this article. In sections \ref{section_wlr} and \ref{samplesizederivationsection} we present the weighted log-rank test and derive an optimal sample size when using this test following \cite{hasegawa2014sample}. This sample size derivation is presented as an alternative to the Schoenfeld's formula \cite{schoenfeld1983sample}, which is normally used when calculating the necessary sample size in confirmatory trials. In section \ref{combinationtest_ch} we introduce the combination test statistic, which will be necessary when we perform sample size re-estimation in adaptive group sequential designs.

Let $T$ be a vector that contains the event times, $t_i, i=1,2, \ldots, D$, between the patients' enrollment date and the patients' final event date, $t_D$,  such that $t_1 < t_2 < \dots < t_D$. Let the number of events at time $t_i$ be denoted as $d_i$, the total number of patients at risk at that time be denoted as $n_i$, and the effect delay (in months) be denoted as $\epsilon$. As previously described if $t < \epsilon$ both survival curves go in parallel and once $t \geq \epsilon$, the survival curves will start diverging. Hence, we assume the following density functions $f_{j}(t)$, survival functions  $S_{j}(t)$ and hazard functions $h_{j}(t)$ for the control group ($j=1$) and for the experimental group ($j=2$):

\begin{equation}
\label{survivalHazards}
\begin{split}
&f_1(t) =\lambda \exp(-\lambda t), ~  S_1(t) = \exp(-\lambda t) \quad \mbox{and} \quad h_1(t) = \lambda, \\
&f_2(t) = \left\{\begin{matrix}\lambda \exp(-\lambda t) \\ c \psi \lambda \exp (-\psi \lambda t) \end{matrix}\right., ~  S_2(t) = \left\{\begin{matrix}\exp(-\lambda t) \\ c \exp (-\psi \lambda t) \end{matrix}\right. \quad \mbox{and} \quad h_2(t) = \left\{\begin{matrix}\lambda & \mbox{if} & 0 \leq t < \epsilon \\ \psi \lambda & \mbox{if} & t \geq \epsilon \end{matrix}\right.,
\end{split}
\end{equation}where $c = \exp \left [ \epsilon \psi \lambda \left ( \frac{1}{\psi -1} \right ) \right ]$ so that $\int_{0}^{\infty} f_2(t) \mbox{d}t = 1$. This way, we assume a step function for the hazard ratio where from time 0 to $\epsilon$, the hazard ratio is equal \com{to} 1, and from time $\epsilon$ the hazard ratio is equal to $1/ \psi$.

In this article we assume that the control group receives the standard of care and the experimental group receives a combination of the standard of care plus the IO agent which causes the delayed effect. Hence, any observed difference from time 0 until time $\epsilon$ is random. The conclusions we obtain are only applicable to studies where a similar assumption is made. Otherwise, we cannot guarantee that from time 0 to time $\epsilon$, both groups have a common survival function.

\subsection{Weighted log-rank test}
\label{section_wlr}

The weighted log-rank test is defined as

\begin{equation}
\label{weightedlogrank}
Z_r = \frac{\sum_{i=1}^{D} r_i (d_{1i} - \mbox{E}(d_{1i}))}{\sqrt{\sum_{i=1}^{D} r_i^2 \mbox{Var}(d_{1i})}},
\end{equation}where $\mbox{E}(d_{1i}) = n_{1i} \times \left ( \frac{d_i}{n_i} \right )$, $\mbox{Var}(d_{1i}) = \frac{n_{1i} n_{2i} d_i (n_i - d_i)}{n_i^2(n_i - 1)}$ and $Z_r \myeq N(0,1)$ under the null hypothesis $H0: h_1/h_2 = 1$.

\cite{fleming1981class} proposed the use of $r_i$ to weight early, middle and late differences through the $G^{\rho,\gamma}$ class of weighted log-rank tests, where the weight function at a time point $t_i$ is equal to

\begin{equation}
\label{weightFunction}
r_i =  \hat{S}(t_i)^\rho (1 - \hat{S}(t_i))^\gamma,
\end{equation}where $\hat{S}(t_i)$ corresponds to the Kaplan-Meier estimator.

Depending on the values of $\rho$ and $\gamma$, we will have different weight functions that will emphasize early differences $(\rho = 1, \gamma = 0)$, middle differences $(\rho = 1, \gamma = 1)$ or late differences $(\rho = 0, \gamma = 1)$ in the hazard rates or the survival curves. The parameter combination attributes equal weights to all $(\rho = 0, \gamma = 0)$ data values and hence does not emphasize any survival differences between treatment arms. Moreover, with this parameter combination \eqref{weightedlogrank} corresponds to the usual log-rank test.

As mentioned by \cite{hasegawa2014sample}, since we focus on the entire survival curve rather than the late difference, valid inference requires pre-specification of $\rho$ and $\gamma$ prior to any data collection.

Prior specification of $(\rho, \gamma)$ is always advisable for the trial integrity, although some authors (see e.g., \cite{lawrence2002strategies}) note that the value of $(\rho, \gamma)$ can be modified at the interim analysis without type-I error rate inflation. At the end of the trial, we are interested in estimating the hazard ratio across the entire study, which is obtained through the standard Cox model \cite{cox1992regression}. \com{Note however that there will be a disconnect between the hazard ratio (i.e., the standard Cox model) and the weighted log-rank test. To obtain an estimate based on the Cox model that corresponds to the weighted log-rank test see \cite{lin2017estimation}.}

In this article we focus on the use of the weighted log-rank test in confirmatory trials with delayed effects. Other areas of use may include treatment switching, which is sometimes present in confirmatory trials and also induces non-proportional hazards (see \cite{bowden2016gaining}). However, it is out the scope of this article to evaluate the performance of the weighted log-rank test under the presence of treatment switching and further research on this matter would be necessary.

\subsection{Sample size derivation for the weighted log-rank test}
\label{samplesizederivationsection}

We introduce the optimal sample size derivation proposed by \cite{hasegawa2014sample}. Assume that we recruit patients during time $T$ at a certain rate in a confirmatory trial where we aim to compare survival time between two groups ($j = 1, 2$): a control group, with a constant hazard over time, and an experimental group, with a hazard that changes over time. The final analysis is performed at time $T + \tau$ after the first patient is enrolled. The study period $[0,T + \tau]$ is partitioned into $M$ subintervals of equal length $\{ t_0 = 0, t_1, t_2, \dots, t_M = T + \tau \}$. Let $h_j(t_i)$ be the hazard function for group $j$ at time $t_i$ and $N_j(t_i)$ be the expected number of patients at risk for group $j$ at time $t_i$, where $i = 0,\dots, M-1$.

\cite{lakatos1988sample} showed that the weighted log-rank statistic is normally distributed with unit variance and approximate expectation of

\begin{equation}
\label{expectation0}
E = \frac{\sum_{i=0}^{M-1} D_i r_i \left ( \frac{\phi_i \theta_i}{1 + \phi_i \theta_i} - \frac{\phi_i}{1 + \phi_i} \right )}{\sqrt{\sum_{i=0}^{M-1} D_i r_i^2 \frac{\phi_i}{(1 + \phi_i)^2}}},
\end{equation}where

\begin{equation}
\begin{split}
& \theta_i = \frac{h_2(t_i)}{h_1(t_i)}, ~~ \phi_i = \frac{N_2(t_i)}{N_1(t_i)}, ~~ D_i = (h_1(t_i) N_1(t_i) + h_2(t_i) N_2(t_i)),\\
& N_j(t_0) = n w_j, ~~ N_j(t_{i+1}) = N_j(t_i) \left [ 1 - h_j(t_i) - \left ( \frac{1}{T + \tau - t_i} \right ) I_{\{ t_i > \tau \}} \right ], \\
\end{split}
\end{equation}$w_j$ represents the allocation ratio for group $j$, and $r_i$ corresponds to the Fleming-Harrington's $G^{\rho,\gamma}$ class of weights where $r_i = (S(t_i))^{\rho}(1 - S(t_i))^{\gamma}$ and $S(t_i)$ represents the pooled survival function. Even though it was originally proposed by \cite{fleming2011counting}, \cite{hasegawa2014sample} uses $S(t_i) = w_1 S_1(t_i) + w_2 S_2(t_i)$ as a substitute for the pooled survival function, where $S_j(t_i)$ represents the survival function of group $j$ at time $t_i$. However, as stated by \cite{hasegawa2014sample}, equation \eqref{expectation0} can be equivalently expressed as

\begin{equation}
\label{expectation}
E = n^{\frac{1}{2}}E^* = n^{\frac{1}{2}} \left [  \frac{\sum_{i=0}^{M-1} D_i^* r_i \left ( \frac{\phi_i \theta_i}{1 + \phi_i \theta_i} - \frac{\phi_i}{1 + \phi_i} \right )}{\sqrt{\sum_{i=0}^{M-1} D_i^* r_i^2 \frac{\phi_i}{(1 + \phi_i)^2}}} \right ],
\end{equation}where

\begin{equation}
\begin{split}
&D_i^* = (h_1(t_i) N_1^*(t_i) + h_2(t_i) N_2^*(t_i)),\\
& N_j^*(t_0) = w_j, ~~ N_j^*(t_{i+1}) = N_j^* (t_i) \left [ 1 - h_j(t_i) - \left ( \frac{1}{T + \tau - t_i} \right ) I_{\{ t_i > \tau \}} \right ], \\
\end{split}
\end{equation}

Assuming that the weighted log-rank statistic is normally distributed with mean $n^{\frac{1}{2}}E^*$ and unit variance, then for a power equal to $1- \beta$ and one-sided significance level $\alpha$ we have

\begin{equation}
\left | n^{\frac{1}{2}}E^*  \right | = z_\alpha + z_\beta,
\end{equation}where $z_\alpha$ and $z_\beta$ correspond to the $\alpha$-th and $\beta$-th percentile of the standard normal distribution respectively. The required sample size is calculated as

\begin{equation}
n = \left ( \frac{z_\alpha + z_\beta}{E^*} \right )^2,
\end{equation}and the total expected number of events is equal to $n \times \sum_{i=0}^{M-1} D_i $.

\subsection{Test statistic}
\label{combinationtest_ch}

We aim to test the null hypothesis, $H_0: \frac{h_1}{h_2} = 1$, against the alternative, $H_1: \frac{h_1}{h_2} < 1$. In the context of group sequential designs, since we are only interested in early efficacy testing we make use of the well known classical group sequential design methodology (see \cite{jennison1999group}) and make use of the O'Brien and Fleming rejection boundaries. In the context of adaptive group sequential designs, we make use of the independent increment property of the inverse normal method, which is an efficient way of incorporating data of patients who where censored at interim analysis while ensuring type-I error rate control (see \cite{wassmer2006planning}). The test statistic is defined as

\begin{equation}
\label{combtest}
Z^* = \com{\xi}_1 \Phi^{-1}(1 - p_1) + \com{\xi}_2 \Phi^{-1}(1 - p_2),
\end{equation} where $p_1$ and $p_2$ denote the separate stage p-values from stages 1 and 2, $\Phi^{-1}$ denotes the inverse of the standard normal distribution, and $\com{\xi}_1$ and $\com{\xi}_2$ are pre-specified weights such that $\com{\xi}_1^2 = \frac{n_1}{n_1 + n_2}$, $\com{\xi}_2^2 = \frac{n_2}{n_1 + n_2}$ and where $n_1$ and $n_2$ represent the number of events observed in each stage. The null hypothesis will be rejected at level $\alpha$ if $Z^* > \Phi^{-1}(1 - \alpha)$.

However, the inverse normal method is in general not valid when doing sample size re-assessment if the adaptations depend on endpoints such OS or PFS (see \cite{bauer2004letter}). We use the approach proposed by \cite{jenkins2011adaptive} where, in equation \eqref{combtest}, the first stage p-value is defined by the cohort of patients included before the interim analysis and is calculated only at the end of the trial. This allows the inclusion of all the events, but it prohibits early stopping for efficacy. See \cite{magirr2016sample} for a detailed review of the existing methods on this matter.


\section{Group sequential and adaptive group sequential designs}
\label{designs_ch}

In this section we aim to briefly describe how group sequential and adaptive group sequential designs work. For a detailed definition and explanation of this methodology see \cite{jennison1999group}.

\subsection{Group sequential designs}

The formulae presented in section \ref{samplesizederivationsection} allow to obtain a sample size that maintains an acceptable power at the end of the trial under the presence of delayed effects. However, a key condition is to have some knowledge about the delay of the drug. Assuming we have this knowledge when designing the confirmatory trial, we can implement a group sequential design with an interim analysis for efficacy. Note that interim analysis for futility is not advised in the presence of delayed effect because of high risk of stopping the study for futility even in scenarios that favor the alternative hypothesis.

\begin{figure}[t]
\caption{Graphical representation of a group sequential design with an interim analysis for efficacy where $\rho_1$ is the efficacy boundary at the interim analysis and $\rho_2$ is the efficacy boundary at the final analysis.}
\centering
\includegraphics[scale=0.6]{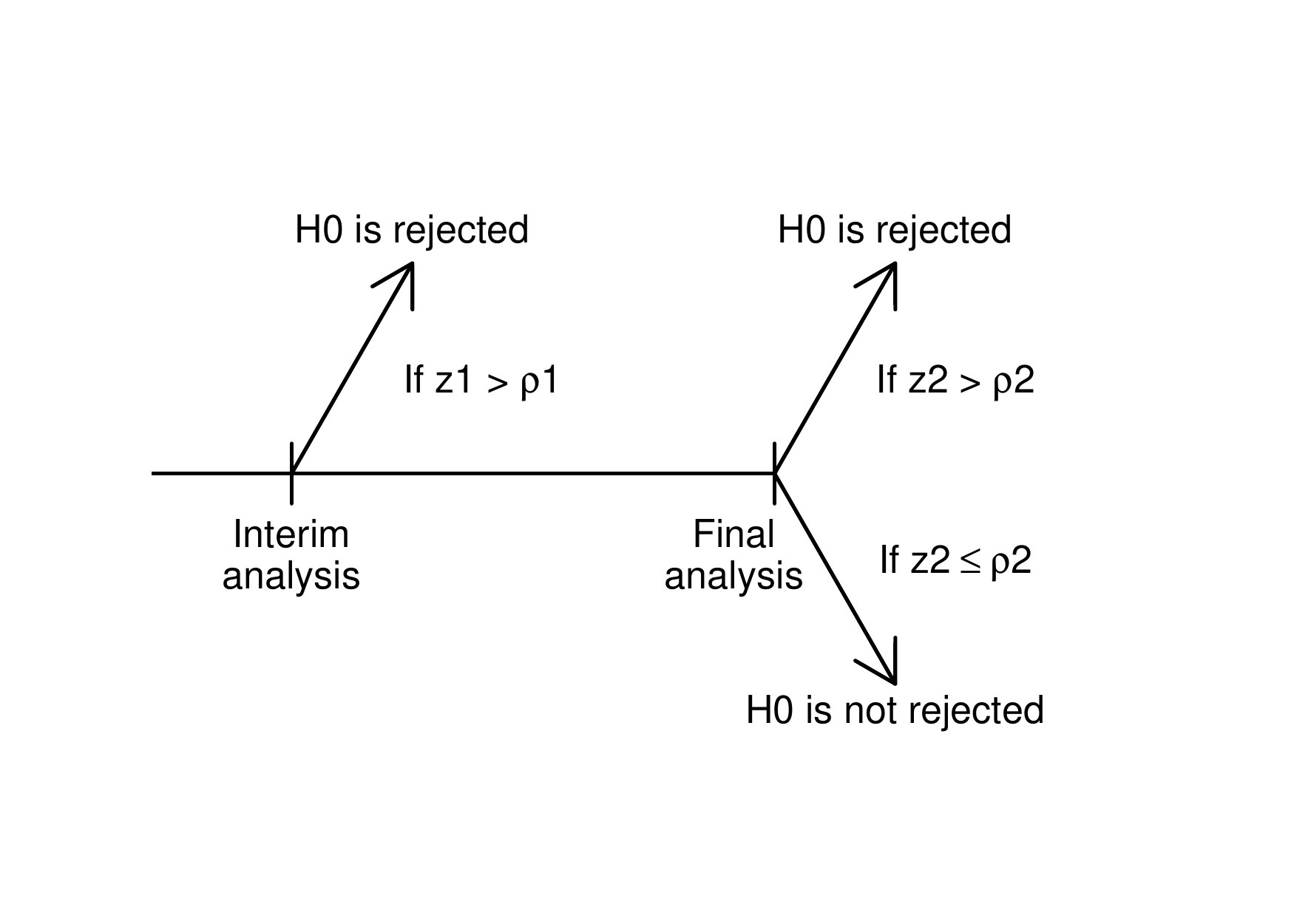}
\label{nonadaptivegsd}
\end{figure}

A group sequential design with one interim analysis for efficacy is graphically described in Figure \ref{nonadaptivegsd}.
  
\subsection{Adaptive group sequential design}

Even though the sample size derivation described in section \ref{samplesizederivationsection} guarantees that after a pre-specified effect delay we will have an acceptable power at the end of the trial while controlling the type-I error rate, we may have misspecified the delay value or maybe this value is unknown. Either way, an adaptive group sequential design that allows interim analyses and sample size re-assessment would be useful in case we expect a lack of statistical power at the end of the trial given the results at the interim analyses. Hence, with this design we aim to recover the power lost due to misspecification of the delay. As explained in section \ref{combinationtest_ch}, to maintain type-I error rate control when the sample size criteria is based on survival endpoints, the interim analysis is only used to do a sample size re-assessment and not for early stopping. Because we need to distinguish between the effect at the interim analysis and the effect at the final analysis, let $\delta_1$ be the hazard ratio at the interim analysis and let $\delta$ be the hazard ratio at the end of the trial.

We now introduce two popular approaches for sample size re-assessment:

\subsubsection{Mehta and Pocock's ``promising zone'' approach \cite{mehta2011adaptive}}
\label{promisingzone_ch}

\cite{mehta2011adaptive} propose a method that adaptively increases the sample size when interim results are considered ``promising''. For that, we compute the conditional power at the interim analysis using $\hat{\delta}_1$ rather than the true $\delta_1$. The formula for the conditional power is defined as

\begin{equation}
\label{conditionalPower}
\mbox{CP}_{\hat{\delta}_1}(z_1,\tilde{n}_2) = 1 - \Phi \left (\frac{z_\alpha \sqrt{n_2} - z_1 \sqrt{n_1}}{\sqrt{\tilde{n}_2}} - \frac{z_1 \sqrt{\tilde{n}_2}}{\sqrt{n_1}} \right ).
\end{equation}

If the conditional power is within a certain pre-specified range that we consider promising, we may re-estimate the sample size to recover the power lost due to the effect delay. The selection of this range depends not only on the estimate of the effect delay but also on the budget of the sponsor for this particular trial. For example, if we have an estimated effect delay between 3 and 7 months, but we only have budget to guarantee 80\% of power up to 5 months, the sponsor can choose to stop the trial. Therefore, following \cite{mehta2011adaptive}, we partition the sample space of attainable $\mbox{CP}_{\hat{\delta}_1}(z_1, \tilde{n}_2)$ values into three zones:

\begin{enumerate}

\item Favorable: We consider the interim results to be in the favorable zone if $\mbox{CP}_{\hat{\delta}_1}(z_1, \tilde{n}_2) \geq 1 - \beta$. In this zone, the study is sufficiently powered for the observed $\hat{\delta}_1$ and therefore no sample size re-estimation is required.

\item Promising: We consider the interim results to be in the promising zone if $1 - \beta > \mbox{CP}_{\hat{\delta}_1}(z_1, \tilde{n}_2) \geq \mbox{CP}_{\min}$. In this zone, $\hat{\delta}_1$ is close to $\delta_1$ but the study is not sufficiently powered and a sample size re-estimation is required. Specifically, the sample size will be increased to

\begin{equation}
\tilde{n}_2^* (z_1) = \min(\tilde{n} \textprime _ 2, n (z_1)_\max),
\end{equation}where $n_\max$ is the maximum sample size the sponsor is willing to enroll and $\tilde{n} \textprime _ 2 (z_1)$ satisfies that $\mbox{CP}_{\hat{\delta}_1}(z_1, \tilde{n} \textprime _ 2) = 1 - \beta$. Following \cite{gao2008sample}, it is possible to show that

\begin{equation}
\tilde{n} \textprime _ 2 = \left ( \frac{n_1}{z_1^2} \right ) \left ( \frac{z_\alpha \sqrt{n_2} - z_1 \sqrt{n_1}}{\sqrt{n_2 - n_1}} + z_\beta \right )^2.
\end{equation}

\item Unfavorable: We consider the interim results to be in the unfavorable zone if the value of $\mbox{CP}_{\hat{\delta}_1}(z_1, \tilde{n}_2) < \mbox{CP}_{\min}$. The value of $\mbox{CP}_{\min}$ is pre-specified before the trial starts and it depends on the prior knowledge about the effect delay. In this zone the interim results are not promising and the sample size will not be re-estimated.

\end{enumerate}

Type-I error rate is controlled following \cite{chen2004increasing}, where it is shown that the overall type-I error does not increases if the sample size is only re-assessed when

\begin{equation}
\label{mehtatype1errorcontrolcriteria}
\mbox{CP}_{\hat{\delta}_1} (z_1) \geq 0.5.
\end{equation}

\subsubsection{Jennison and Turnbull's ``start small then ask for more'' approach \cite{jennison2015adaptive}}
\label{jennison_ch}

\cite{jennison2015adaptive} made a detailed analysis of Mehta and Pocock's ``promising zone'' approach.

One drawback of the ``promising zone'' approach is the use of $\hat{\delta}_1$ in the construction of the promising zone and sample size increase function. The reason is that $\hat{\delta}_1$ is considered as a highly variable estimate of $\delta_1$, and also because it is used twice in determining the conditional power that underlies the sample size function: the first time through the value of $z_1$ and the second time when evaluating the conditional power at $\delta = \hat{\delta}_1$. This double use of $\hat{\delta}_1$ was also pointed out by \cite{glimm2012comments} who recommends a careful inspection of the operating characteristics when using $\delta = \hat{\delta}_1$.

Another drawback of Mehta and Pocock's ``promising zone'' approach is that, despite the type-I error rate being controlled, because of the restriction showed in \eqref{mehtatype1errorcontrolcriteria}, the gain in power is relatively small for the increases in the expected sample size. Moreover, \cite{jennison2015adaptive} demonstrated that other alternatives such us a fixed sample design and a group sequential design have exactly the same power curve and a lower expected sample size around the true value of $\delta$.

To overcome the last limitation, \cite{jennison2015adaptive} propose an optimal sample size calculation rule where we need to find the value of $n_2^*$ that maximizes the objective function

\begin{equation}
\label{objectivefunction}
f(n_2^*) = \mbox{CP}_{\hat{\delta}_1}(z_1, n_2^*) - \eta(n_2^* - n_2),
\end{equation}where $\eta$ can be considered as ``a tuning parameter that controls the degree to which the sample size may be increased when interim data are promising but not overwhelming''.

\cite{jennison2015adaptive} pointed out that even though the objective function given by equation \eqref{objectivefunction} ``concerns conditional probabilities given the interim data, choosing a sample size rule to optimize this objective function also yields a design with an overall optimality property expressed in terms of unconditional power''. They show that

\begin{equation}
\label{unconditionalpower}
P_{\hat{\delta}_1}(\mbox{Reject} ~ H_0) - \eta E_{\hat{\delta}}(N) = \int \{ \mbox{CP}_{\hat{\delta}}(z_1, n_2^*(z_1)) - \eta(n_2^*(z_1) - n_2) \} f_{\hat{\delta}}(z_1) d z_1,
\end{equation}where $f_{\hat{\delta}}(z_1)$ represents the density of $Z_1$ under $\delta = \hat{\delta}$, and since we maximize equation \eqref{objectivefunction}, for every $z_1$, we also maximize the right hand side of equation \eqref{unconditionalpower}. Moreover, it is possible to show that this sample size rule has the minimum expected sample size among all rules that achieve the sample power under  $\delta = \hat{\delta}$.

In algorithms \ref{IOAdaptiveDesign_algorithm} and \ref{IOAdaptiveDesign_algorithm_jennison} we describe how to implement the reviewed methodology in case the sample size needs to be re-assessed.

\begin{algorithm}[H]
  \caption{Group sequential adaptive design using Mehta and Pocock's ``promising zone'' approach.}
  \label{IOAdaptiveDesign_algorithm}
  \begin{algorithmic}[1]
    \Procedure{}{}

      \State Recruit up to $n$ patients and when $n_1$ events are observed analysis compute CP$_{\hat{\delta}_1}(z_1,\tilde{n}_2)$
      \State Calculate the number of events $\tilde{n} \textprime _ 2$ and total sample size necessaries for the second stage.
      \State Recruit patients until $\tilde{n} \textprime _ 2$ events are observed.
\State Compute  $Z^* = \com{\xi}_1 \Phi^{-1}(1 - p_1) + \com{\xi}_2 \Phi^{-1}(1 - p_2)$ \Comment{$p_1$ is calculated at the final stage using only the patients enrolled before the interim analysis.}
 \If{($Z^* > Z_\alpha$)}
  	
  	  \State Outcome $\gets 1$ \Comment{$H_0$ is rejected at the final analysis}
  	
  	 \Else
  	
  	  \State Outcome $\gets 0$ \Comment{$H_0$ is not rejected at the final analysis}
  	
  	 \EndIf

    \State \textbf{return} Outcome
    \EndProcedure
  \end{algorithmic}
\end{algorithm}

\begin{algorithm}[H]
  \caption{Group sequential adaptive design with one interim analysis for efficacy using Jennison and Turnbull's ``start small then ask for more'' approach.}
  \label{IOAdaptiveDesign_algorithm_jennison}
  \begin{algorithmic}[1]
    \Procedure{}{}

      \State Recruit up to $n$ patients, and when $n_1$ events are observed do the interim analysis.
       	
  	\State Calculate the number of events $n_ 2^*$ and total sample size necessaries for the second stage.
  	
  	\State Recruit patients until $n_2^*$ events are observed.
  	
  	 \State Compute  $Z^* = \com{\xi}_1 \Phi^{-1}(1 - p_1) + \com{\xi}_2 \Phi^{-1}(1 - p_2)$ \Comment{$p_1$ is calculated at the final stage using only the patients enrolled before the interim analysis.}

  	  \If{($Z^* > Z_\alpha$)}
  	
  	  \State Outcome $\gets 1$ \Comment{$H_0$ is rejected at the final analysis}
  	
  	 \Else
  	
  	  \State Outcome $\gets 0$ \Comment{$H_0$ is not rejected at the final analysis}
  	  	  	
  	\EndIf

    \State \textbf{return} Outcome
    \EndProcedure
  \end{algorithmic}
\end{algorithm}


\section{Simulation setup}
\label{section_simulation}

We implement the methodology described in sections \ref{sectionMethods} and \ref{designs_ch} on a scenario that tries to imitate a realistic phase III trial with delayed effects in oncology.

Survival data for the control arm is simulated using an exponential distribution while data for the experimental arm is simulated using a distribution that is piece-wise exponential (see equation \eqref{survivalHazards}). Under proportional hazards, we assume that the control arm has a median survival of 6 months while the experimental arm has a median survival of 9 months. Hence, the hazard ratio is equal to 0.667. However, under the presence of delayed effects we assume a step function for the hazard ratio where it will be equal to 1 until a certain time point $\epsilon$, and then it will be at its full effect after $\epsilon$. This means that while the control arm will keep its median survival of 6 months, the median survival of the experimental arm will no longer be 9 months because of the delayed effect.

We establish a total study duration of 25 months, a total enrollment period of 17.5 months, randomization ratio 1:1, a power of 90\% and a one-sided level $\alpha$ of 2.5\%.

Clinical trial enrollment follows a Poisson distribution with rate of 10 patients per month. Plotting the cumulative distribution function of a Poisson distribution of these characteristics using, for instance, the R function \texttt{ecdf()}, it is straightforward to see that after 17.5 months almost all the patients, if not all, are enrolled in the trial. Results are obtained running 200,000 simulated trials. R code is showed in the appendix explaining how to simulate survival data under the presence of delayed effects.

In Table \ref{table_efficacyBoundaries} we show the information fraction, the cumulative $\alpha$ spent, the O'Brien and Fleming efficacy boundaries, and the boundary crossing probability at each look. Recall that these boundaries are only used in the context of group sequential designs where the sample size is not re-assessed and they are calculated based on the information fraction only. If the sample size needs to be re-assessed, we employ different methodology (see section \ref{combinationtest_ch})

\begin{table}[H]
\centering
\caption{Information fraction, the cumulative $\alpha$ spent, the efficacy boundaries, and the boundary crossing probability at each analysis in the group sequential design we use as an example.}
\label{table_efficacyBoundaries}
\begin{tabular}{|c|c|c|c|c|c|}
\hline
Look \# & \begin{tabular}[c]{@{}c@{}}Information \\ Fraction\end{tabular} &   \begin{tabular}[c]{@{}c@{}}Cumulative $\alpha$ \\ spent \end{tabular}  & \begin{tabular}[c]{@{}c@{}}Efficacy \\ boundary Z \end{tabular}  & \begin{tabular}[c]{@{}c@{}}Boundary crossing \\ probability (incremental) \end{tabular} \\ \hline
1       & 0.75                 &      0.01                      & 2.34   & 0.688   \\ \hline
2       & 1                    &     0.025                     & 2.012  & 0.212   \\ \hline
\end{tabular}
\end{table}

For both the group sequential and the adaptive group sequential designs, we estimate the empirical power and the empirical type-I error rate at the final analysis. In the context of group sequential designs, let Z$_{\tiny \mbox{test}}$ be the Z-statistic obtained at the end of the trial and Z$_2$ be the efficacy boundary of the final analysis presented in Table \ref{table_efficacyBoundaries}. In scenarios under the alternative hypothesis, the empirical power is defined as

\begin{equation}
\label{powerequation}
\mbox{Power} = \frac{1}{M} \sum_{i=1}^{m} I[\mbox{Z}_{\tiny \mbox{test}} > \mbox{Z}_2],
\end{equation}whereas in scenarios under the null hypothesis, \eqref{powerequation} is the empirical type-I error rate. In the context of group sequential adaptive designs, in equation \eqref{powerequation}, Z$_{\tiny \mbox{test}}$ needs to be substituted by Z$^*$ and Z$_2$ needs to be substituted by Z$_\alpha$ in order the implement the inverse normal method described in section \ref{combinationtest_ch}.

\section{Results}

In this section we evaluate the repercussion of delayed effects on the power and the type-I error rate in group sequential and adaptive group sequential designs. The results presented in this section are based on the simulated scenario described in section \ref{section_simulation}.

Because one of the purposes of this work is to make a comparison between the log-rank test and weighted log-rank test, in Table \ref{table_events} we show, for different delay times, the required number of events and the sample size using the parameter values $(\rho = 0, \gamma = 0)$ and $(\rho = 0, \gamma = 1)$ following the formulas presented in section \ref{samplesizederivationsection}. As we can see, under proportional hazards the parameter combination $(\rho = 0, \gamma = 0)$ is more efficient since it requires 258 events whereas the parameter combination $(\rho = 0, \gamma = 1)$ requires 369 events to maintain 90\% of power. However, with 5 months delay, the parameter combination $(\rho = 0, \gamma = 1)$ becomes more efficient since it requires 741 events whereas the parameter combination $(\rho = 0, \gamma = 0)$ requires 1436 events to maintain 90\% of power.

\begin{table}[H]
\centering
\caption{Sample size calculation for different effect delay times using the parameter values $(\rho = 0, \gamma = 0)$ and $(\rho = 0, \gamma = 1)$ using the sample size formulae reviewed in Section  \ref{samplesizederivationsection}.}
\label{table_events}
\begin{tabular}{|c|c|c|c|c|c|c|c|}
\hline
                                          & Delay (months) & 0   & 1   & 2   & 3   & 4    & 5    \\ \hline
\multirow{2}{*}{$(\rho = 0, \gamma = 0)$} & \# of events   & 258 & 359 & 492 & 686 & 986  & 1436 \\ \cline{2-8}
                                          & \# of patients & 330 & 456 & 621 & 860 & 1228 & 1777 \\ \hline
\multirow{2}{*}{$(\rho = 0, \gamma = 1)$} & \# of events   & 369 & 376 & 406 & 468 & 578  & 741  \\ \cline{2-8}
                                          & \# of patients & 472 & 478 & 512 & 587 & 719  & 917  \\ \hline
\end{tabular}
\end{table}

\subsection{Group sequential design}

\begin{figure}[H]
\caption{Empirical power and type-I error for a wide range of combinations of $\rho$ and $\gamma$ at the final analyses with different effect delay times and a unique sample size calculated assuming proportional hazards. In black, the five combinations with less cumulative power loss over time, in dark grey the power loss of the log-rank test ($\rho=0, \gamma=0$) over time, and in light grey the power loss of the rest of the combinations.}
\centering
\includegraphics[scale=0.5]{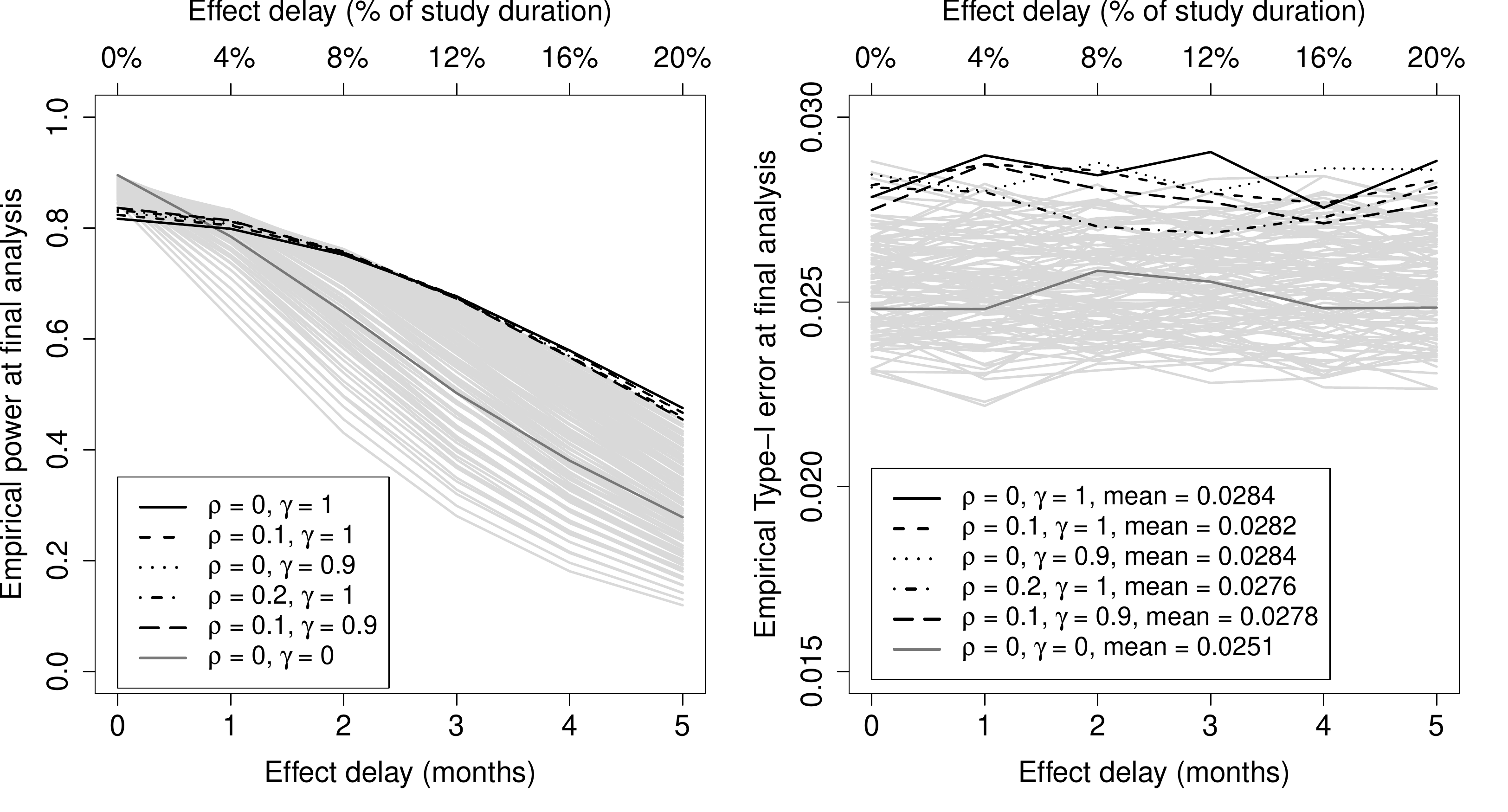}
\label{power_allCombinations_t0}
\end{figure}

In Figure \ref{power_allCombinations_t0} we show the empirical power and type-I error rate at the final analysis for a wide range of $\rho$ and $\gamma$ combinations with the design characteristics presented in section \ref{section_simulation} assuming no delayed effect in the sample size calculation. As expected, the results show that the parameter combination $(\rho = 0, \gamma = 0)$ achieves 90\% of power and 2.5\% type-I error at the final analysis. However, as the delay increases, we observe that power drops faster than other combinations of $\rho$ and $\gamma$ as the effect delay increases. Other combinations like $(\rho = 0, \gamma = 1)$ have less power under proportional hazards but maintain higher power as the effect delay increases. These results are expected since low values of $\rho$ and high values of $\gamma$ weight late differences, which is the situation we recreate in this simulated trial. However, combinations that weight late differences produce a slight type-I error rate inflation as we can observe in Figure \ref{power_allCombinations_t0}, right image.

\begin{figure}[H]
\caption{Empirical power (left) and type-I error (right) for a wide range of combinations of $\rho$ and $\gamma$ at the final analyses with different effect delay times and a different sample size for each delay time. In the left image, in black, the five combinations with highest mean power over time. In dark grey the log-rank combination ($\rho=0, \gamma=0$) and in light grey the rest of the combinations. In the right image, in black the type-I error of the five combinations with highest mean power over time. In dark grey the log-rank combination ($\rho=0, \gamma=0$) and in light grey the rest of the combinations.}
\centering
\includegraphics[scale=0.5]{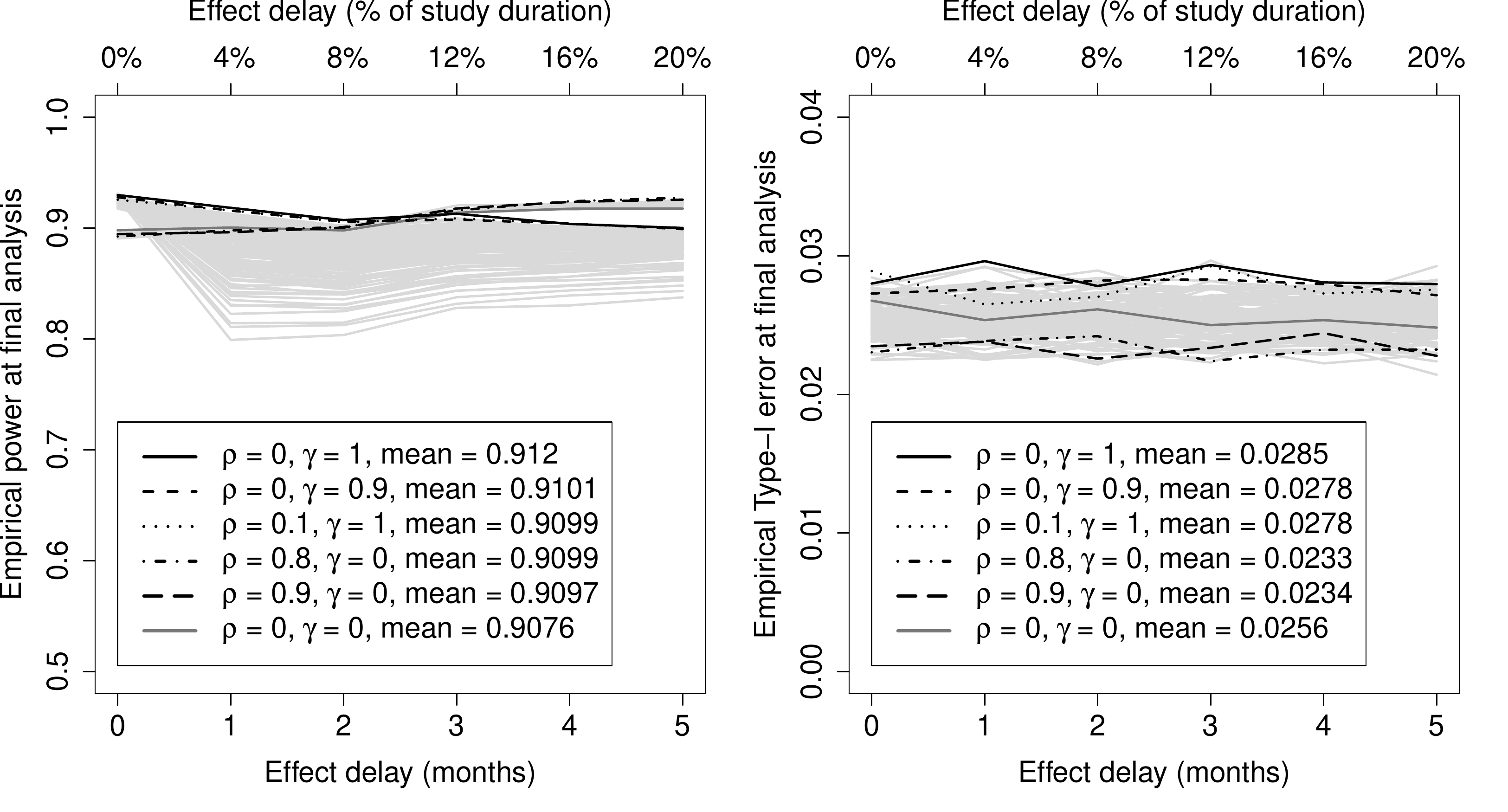}
\label{power_allCombinations_correct_t}
\end{figure}

Using the methodology described in section \ref{samplesizederivationsection}, if we incorporate an estimate of the effect delay in the sample size calculation, we are able prevent the power to drop until that specified moment. This is shown in Figure \ref{power_allCombinations_correct_t}, where for each delay time we calculate the sample size necessary to achieve 90\% power taking the delay into account. Moreover, when correctly specifying the effect delay, we observe that not only low values of $\rho$ and high values of $\gamma$ achieve high power. However, in terms of type-I error rate, we observe the same slight type-I error rate inflation we observed in Figure \ref{power_allCombinations_t0} for low values of $\rho$ and high values of $\gamma$.

To control the type-I error rate, we propose to use a similar approach as the one used by \cite{golkowski2014blinded} in which, although in a different context, instead of calculating the sample size for $\alpha = 2.5\%$, a lower value of $\alpha$ is fixed so the final type-I error rate is maintained at 2.5\%.

\subsection{Adaptive group sequential design}

In this section we show how performing a sample size reassessment we recover some of the power lost due to the delayed effect. As in the previous section, the results presented here make use of the simulated example described in section \ref{section_simulation}. However, rather than using a wide range of combinations of $\rho$ and $\gamma$, we use the combination $(\rho=0,\gamma=1)$ since we believe it is the most suitable combination for this kind of setting.

\begin{figure}[H]
\caption{Empirical type-I error rate (top left), empirical power (top right), percent of times sample size is reassessed (bottom left) and ratio between the reassessed number of events and the original number of events (bottom right) at different delay times, when the sample size is calculated assuming no delay, using the ``promising zone'' approach.}
\centering
\includegraphics[scale=0.5]{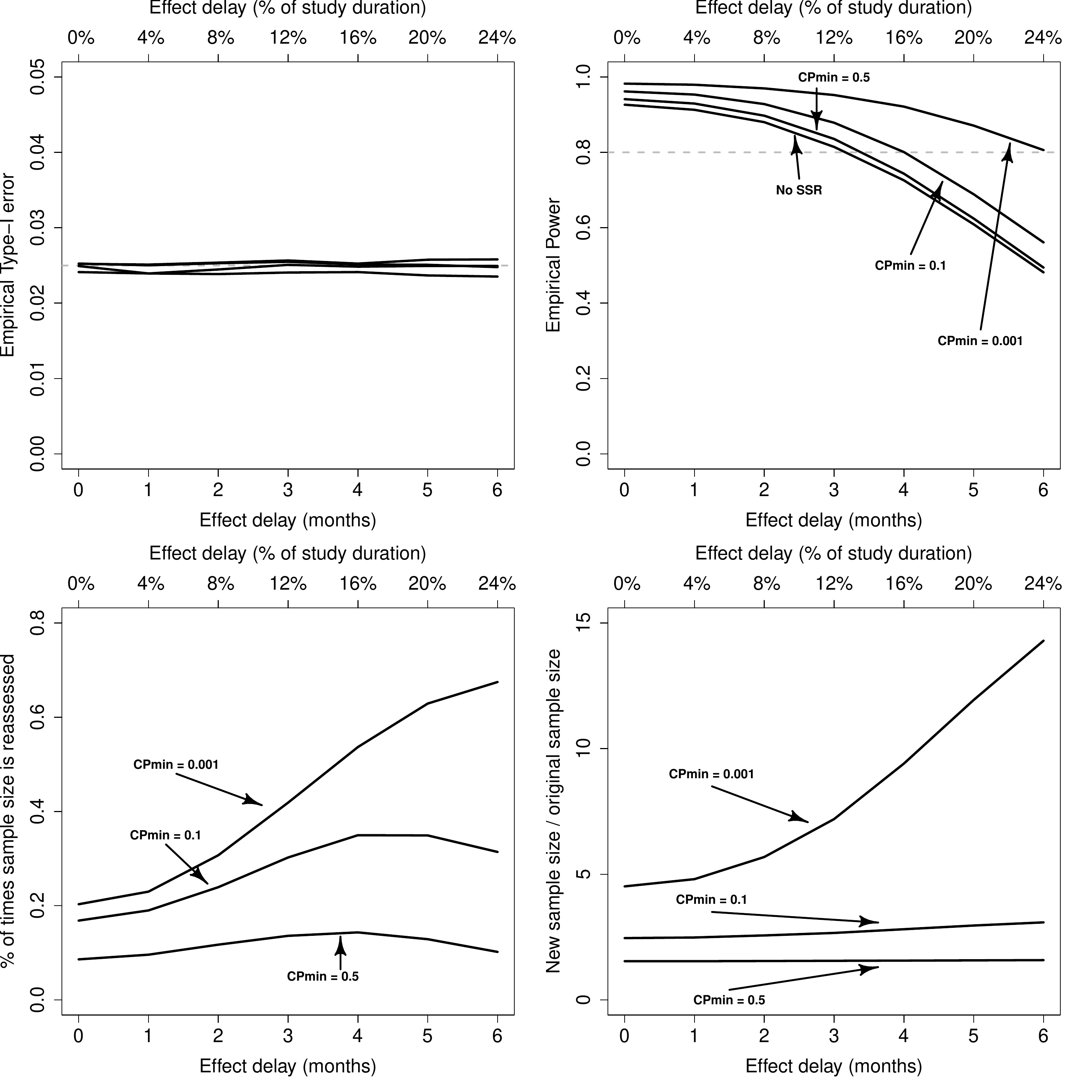}
\label{allPlotsCP}
\end{figure}

In Figure \ref{allPlotsCP} we present the empirical type-I error (top-left image), empirical power (top-right image), percent of times we re-adjust the sample size (bottom-left image) and the ratio between new sample size and original sample size (bottom-right image) for different effect delays using the weighted log-rank test with the parameter combination $(\rho = 0, \gamma = 1)$ using the promising zone approach proposed by \cite{mehta2011adaptive}.

We employ three different promising zone lower bounds (0.5, 0.1, 0.001) and compare their operating characteristics against a design that does not reassess the sample size. Without any sample size reassessment, the power is below 80\% after 3 months. Using a promising zone lower bound of 0.5, the power will be below 80\% after 3.5 months. However, if the promising zone lower bound is 0.1 or 0.001, the power will be below 80\% after 4 and 6 months, respectively. As discussed in the literature (see \cite{jennison2015adaptive}) we corroborate that the gains when using a lower bound of 0.5 is practically negligible and the greatest gains in power are likely to be found outside the region defined by \cite{mehta2011adaptive}.

In terms of type-I error, we observe it is perfectly controlled for any value of the promising zone lower bound. However, note that we implemented our previously described proposal in which instead of calculating the sample size for $\alpha = 2.5\%$, a lower value of $\alpha$ is fixed so the final type-I error rate is maintained at 2.5\%. Otherwise we would see the same slight type-I error rate inflation we identified in the Figures \ref{power_allCombinations_t0} and \ref{power_allCombinations_correct_t} due to the $\rho$ and $\gamma$ parameters that we employ.

In terms of percent of times we fall in the promising zone, when the lower bound is 0.5, the probability of re-adjusting the sample size reaches its maximum value, which is around 15\% at 4 months. If the lower bound is 0.1, the probability of re-adjusting the sample size reaches its maximum value, which is around 35\% between 4 and 5 months. Last, if the lower bound is 0.001, the probability of re-adjusting the sample size reaches its maximum value, which is close to 70\% at 6 months.

In terms of how much we need to increase the sample size with respect to the original sample size every time we fall in the promising zone, we observe that if the lower bound is 0.5, we need around 1.5 times the original sample size regardless the delay time. If the lower bound is 0.1, we need around 2.5 times the original sample size also regardless the delay time. Last, if the lower bound is 0.001, for a delay time $t=0$, we need around 4.5 times the original sample size. For a delay time $t=4$ we need around 9 times the original sample size and for a delay time $t=6$ we need around 15 times the original sample size.

It is important to mention that, in practice, a promising zone lower bound of 0.001 may not be possible to implement given the excessively increase in the number of events needed and the consequent increase in the budget for the trial. However, we believe it is interesting to show that it is possible to maintain a power of 80\% for another three extra months, regardless of the additional duration and expenses of the trial.

\begin{figure}[H]
\caption{Empirical power and ratio between the reassessed number of events and the original number of events when using the approaches from \cite{mehta2011adaptive} and \cite{jennison2015adaptive}}
\centering
\includegraphics[scale=0.5]{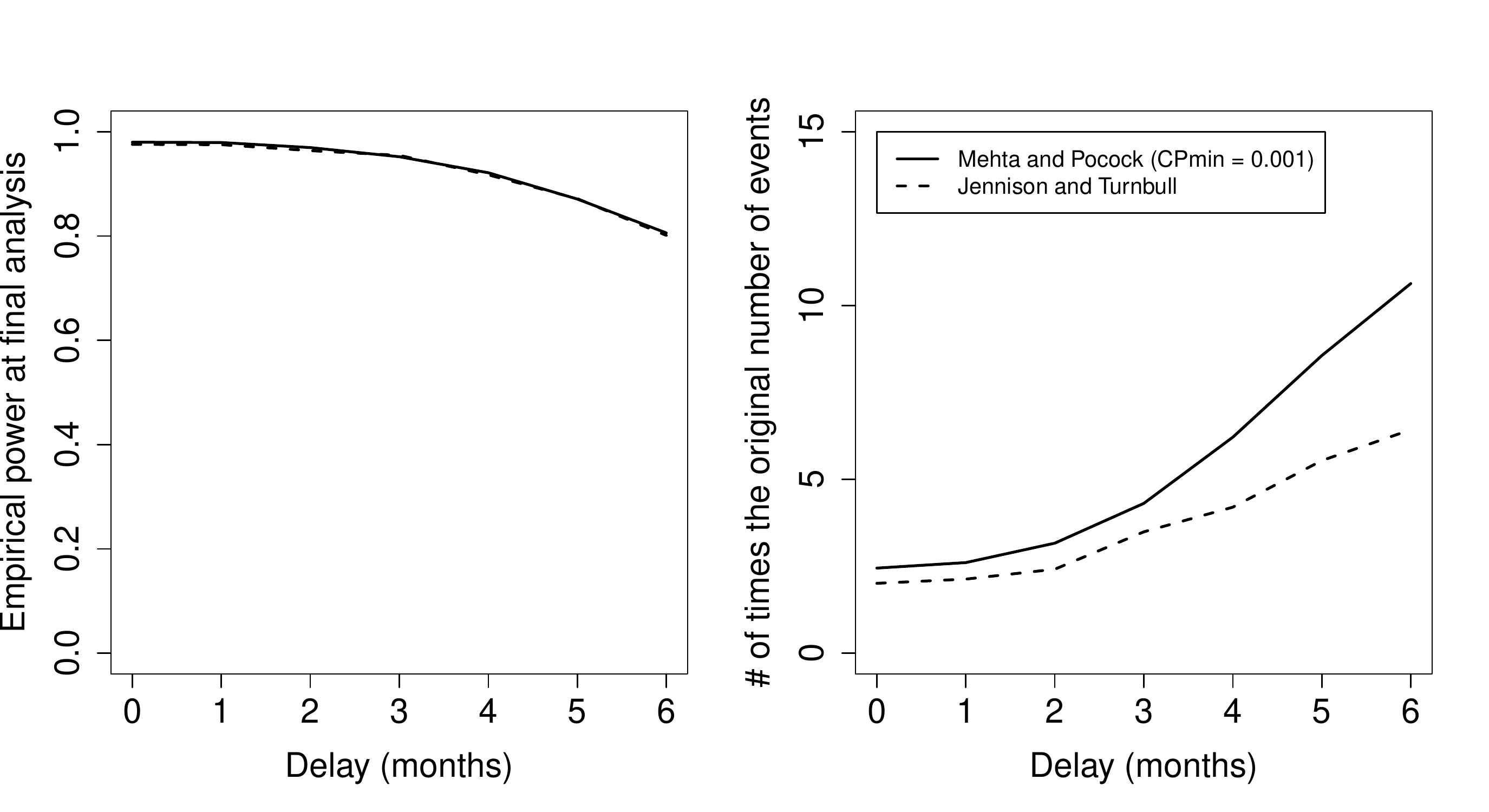}
\label{mehtaVSjennison}
\end{figure}

Last, in Figure \ref{mehtaVSjennison} we make a comparison between the approaches of \cite{mehta2011adaptive} and \cite{jennison2015adaptive}. We selected the promising zone's lower bound 0.001 because it is the one that is more expensive to put into practice and where greater differences are observed. As expected, the approach from \cite{jennison2015adaptive} is able to maintain the same power as the approach from \cite{mehta2011adaptive}. However, in terms of how much we need to increase the sample size with respect to the original sample size, \cite{jennison2015adaptive} requires smaller sample size, specially after 4 months of delay.

\section{Practical Considerations}

In the previous sections we evaluated the impact of delayed effects in clinical trials and what methodology exists in order to reduce it. However, we cannot conclude which methodology is better in general terms because it will depend on many factors. In this section, we emphasize some practical considerations regarding the use of the presented methodology.

The first question we tackled in this article in the use of the weighted log-rank test versus the log-rank test in group sequential and adaptive group sequential designs. In the presence of known delayed effects, we observed that the weighted log-rank test with parameter values $(\rho=0, \gamma=1)$ is the overall best choice, not only for the analysis but also for the sample size formula. We recall that the use of these parameter values in the weighted log-rank test generates a slight type-I error rate inflation and hence the value of $\alpha$ needs to be slightly decreased in order to achieve a final type-I error rate of 2.5\%.

In cases where the delayed effect is unknown or underestimated in the sample size calculation, there exists methodology that re-adjusts the sample size in order to increase the power at the final analysis. The use of each method will depend on the characteristics of the trial. From the two methods we evaluated in the article, we observed that the proposal of \cite{jennison2015adaptive} outperforms the proposal of \cite{mehta2011adaptive} in the sense that for the same power, \cite{jennison2015adaptive} requires less sample size. However, with these approaches it is possible to back-calculate the conditional power at the interim analysis if we know the sample size increase recommended for the second stage of the trial. If this situation does not compromise the integrity of the trial, we recommend the use of \cite{jennison2015adaptive} as it is proven to be more efficient. However, if the effect at the interim analysis has to remain masked, we propose the use of a modified version of \cite{mehta2011adaptive}, which would work as follows.

\begin{figure}[H]
\caption{Fixed sample size increase illustration following a modified version of the ``promising zone'' proposed by \cite{mehta2011adaptive}.}
\centering
\includegraphics[scale=0.5]{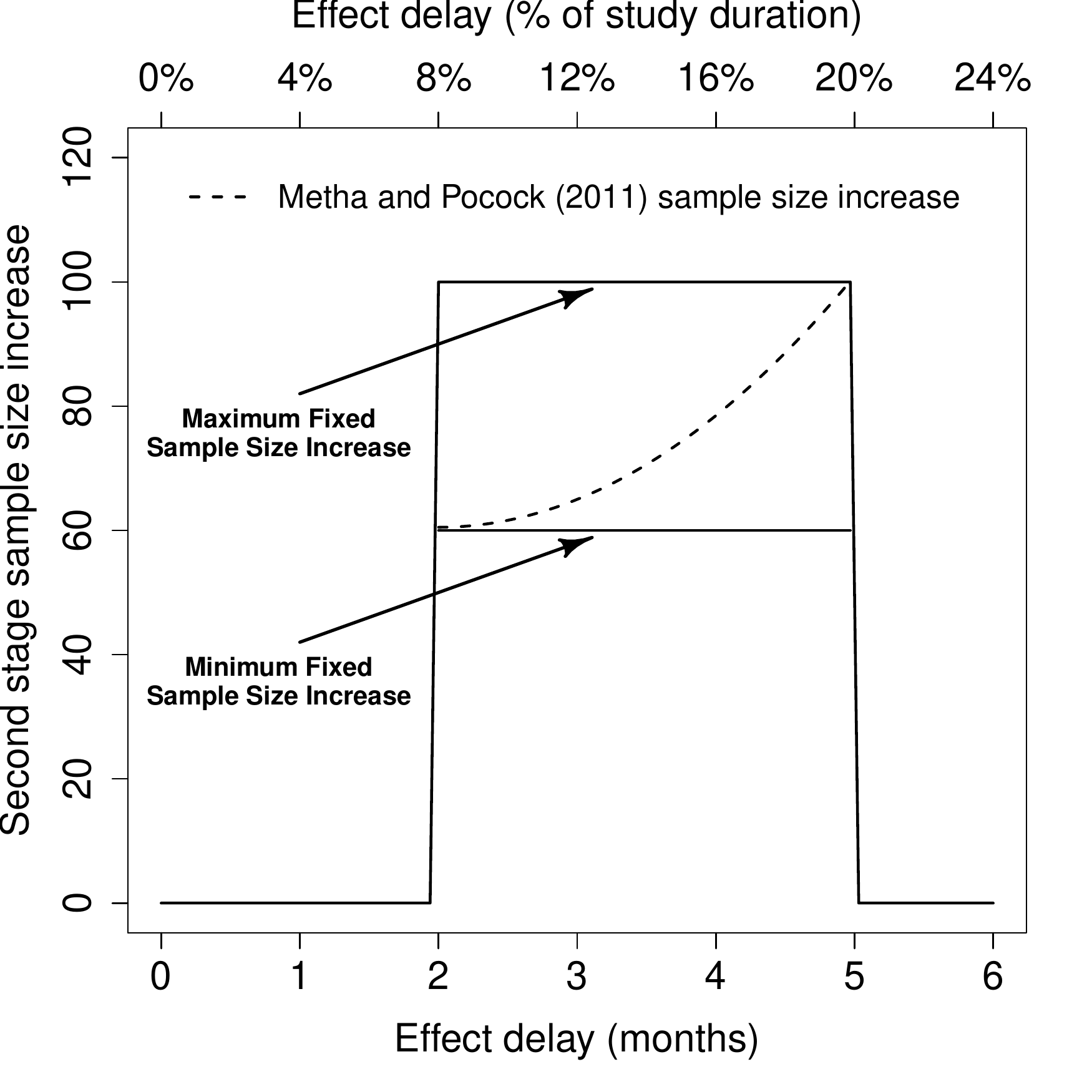}
\label{fixed_sample_size_increase}
\end{figure}

We would establish a promising zone, as in the original method, in which we re-calculate the sample size if the conditional power falls within a certain pre-specified range. The original method would calculate a different sample size for each conditional power (or delay time). However, in order to avoid back-calculations based on the second stage sample size, we propose to fix in advance the sample size to be used in the second stage of the trial. To avoid having an underpowered trial, we can fix the sample size increase assuming the lowest possible value for the conditional power (or the highest delay time) of the promising zone. This value would represent the maximum fixed sample size increase, although with this approach, we will be overpowering the trial for almost all values of the conditional power that fall in the promising zone. On the other hand, we can also fix the sample size increase to the highest possible value for the conditional power (or the lowest delay time) of the promising zone. This fixed value would represent the minimum fixed sample size increase, although with this approach, we will be underpowering the trial for almost all values of the conditional power that fall in the promising zone. This modification of the method proposed by \cite{mehta2011adaptive} is illustrated (using a toy example) in Figure \ref{fixed_sample_size_increase}. 

In this case, even though the ``safest'' option would always be using the maximum fixed sample size increase, we cannot give a recommendation since a large number of sample sizes between the maximum and the minimum fixed sample size increases can be employed and the choice depends on how much risk of having an underpowered study the sponsor is willing to take.

\section{Conclusions}

In this article we evaluated the impact of delayed effects, in terms of power and type-I error rate, in phase III clinical trials. We studied the use of the weighted log-rank test as an alternative to the log-rank test in group sequential and adaptive group sequential designs. This includes not only the analysis but also the incorporation of the Fleming and Harrington class of weights, as well as a delay estimate, in the sample size calculations. Also, we reviewed two different sample size re-adjustment methods, and explored which one is more efficient.

Results show that, in the presence of delayed effects when assuming proportional hazards, the weighted log-rank test with parameter values $(\rho=0, \gamma=1)$ was the best overall choice, as it was the one that maintained a higher power as the delay increases. When incorporating the Fleming and Harrington class of weights, as well as a delay estimate, into the sample size calculation, we observed that the power remains until the delay estimate we provided and the difference in terms of power between parameter values was not as big as under the assumption of proportional hazards, although the parameter values $(\rho=0, \gamma=1)$ were overall the best combination. Sample size re-adjustment allows increasing the sample size at the interim analysis to lower the risk of failing to meet the study objective. We explored the operating characteristics of two popular approaches for sample size re-adjustment: the ``promising zone'' approach by \cite{mehta2011adaptive} and the ``start small then ask for more'' approach by \cite{jennison2015adaptive}.

With the proposal from \cite{mehta2011adaptive} it is possible to maintain the power high enough for the trial to be valid. However, the proposal from \cite{jennison2015adaptive} is proven to be more efficient as for the same power curve, it requires less sample size. Nevertheless, there are situations in which having a ``promising zone'' may be more beneficial. This is the case when the effect at the interim analysis has to remain masked for integrity reasons. The problem is that it is possible to back-calculate the effect at the interim analysis by knowing the sample size increase. Hence, in this article we propose a modified version of the proposal from \cite{mehta2011adaptive}. It does not require any modification of the original formulation. If a trial has a conditional power that falls in a pre-specified promising zone, we apply a pre-specified fixed sample size increase that will be used regardless the value of the conditional power as long as it falls in the promising zone. With this approach, even though we maintain the effect masked at the interim analysis, there is the risk of having an underpowered study if the fixed sample size increase in not large enough. However, if we want to avoid that risk, we will need to recruit more patients than necessary with the associated extra cost.

\section{Acknowledgments}
We would like to acknowledge Michael Branson for inspiration and initial discussions, and Ekkehard Glimm, Franz K\"{o}nig and Thomas Jaki for the valuable advice during the development of this work. This project has received funding from the European Union's Horizon 2020 research and innovation programme under the Marie Sklodowska-Curie grant agreement No 633567.


\bibliography{references.bib}
\bibliographystyle{pharmstat3} 

\newpage{}

\section*{Appendix}

In this section of the manuscript we present the R code we used to simulate survival data for confirmatory and for running a group sequential design with one interim analysis for efficacy under the presence of delayed effects.

\begin{lstlisting}[language=R]

#Sample size
n = 330

#Overall number of events
nevents = 258

#Number of events at the interim analysis
nevents_look1 = 194

#Enrollment rate
enrollment_rate = 10

#Survival median of the control group 
originalMedian1 = 6

#Survival median of the experimental group
originalMedian2 = 9

#Delay
epsilon = 2

#Parameters rho and gamma (remember that rho=0 and 
#gamma=0 implies using the usual log-rank test).
prho = 0
pgamma = 0

#Interim analysis rejection boundary
rejectionBoundary1 = 2.339711

#Final analysis rejection boundary
rejectionBoundary2 = 2.011719

#Simulated trial number 'k' (if you want to compute the power 
#and type-I error of this design a for loop is needed since this code only 
#simulates one clinical trial).
k = 1
        
#We generate the survival times for all the patients. Note that because until
#the delayed effect kicks the hazard ratio is equal to one, and hence all the
#survival data is generated from the same exponential distribution.
event_time = rexp(n, rate = log(2)/originalMedian1)

#We generate the treatment arm.
group = c(rep(0, n/2),rep(1,n/2))
        
#We generate the enrollment times.
enrollment_times = rpois(n,enrollment_rate)

#We select the treatment arm with immuno-therapy and if the survival
#time is larger than the moment where the delay kicks in (epsilon), 
#we resample that value.
for(i in 1:n){
  if(group[i] == 1 & event_time[i] > epsilon){
    event_time[i] =  epsilon + rexp(1, rate = log(2)/originalMedian2)
    }
}
        
#We create a variable containing what we call pseudo-survival times. 
#It contains the sum of the enrollment time plus the survival time and 
#we use it to establish the censoring threshold using the number of events
#required to do the interim analysis.
pseudo_surv = enrollment_times+event_time
        
#INTERIM ANALYSIS

#We sort our pseudo-survival variable and choose the time of 
#the observation number that is equal to the number of events 
#required to do the interim analysis.
threshold_censoring[k,1] = sort(pseudo_surv)[nevents_look1]

#We create a variable to identify enrolled patients.
enrolled = ifelse(enrollment_times < threshold_censoring[k,1], 1, 0)

#We create a variable that contains the survival times taking 
#into account the censoring status.
survival_time = ifelse(pseudo_surv > threshold_censoring[k,1],
threshold_censoring[k,1] - enrollment_times, event_time)
 
#We create a variable with the censoring status
censor = ifelse(pseudo_surv > threshold_censoring[k,1], 0, 1)

#We put all the variables together and keep only the enrolled 
#patients for the analysis.
data.df1 <- data.frame(survival_time,censor,group,enrolled)        
data.df1 = data.df1[data.df1[,4]==1,]

#We do the weighted log-rank test to reject or not the null hypothesis.
fit.wlr00_st1 <- wtdlogrank(Surv(survival_time, censor)~group, 
data = data.df1, sided = 1, WtFun="FH", param=c(prho,pgamma))

#If z1 > rho1 (according to manuscript notation) we reject the 
#null hypothesis that states that the hazard ratio is equal to one.
reject[k,1] = ifelse(abs(fit.wlr00_st1$Z) > rejectionBoundary1, 1, 0)

#This is only to be sure that we account for the rejection at the interim 
#of the null hypothesis in the final analysis as well, even though we 
#don't technically do the final analysis because the hypothesis is 
#rejected at the interim and the trial stopped.
if(reject[k,1] == 1){
  reject[k,2] = 1
  next
}

#We now continue with the second stage of the trial if 
#we didn't stop for efficacy.

#For the second stage we repeat the same procedure we did for the first stage.
#However, this time we do the calculations with the censoring threshold 
#obtained when the total number of events is reached.

threshold_censoring[k,2] = sort(pseudo_surv)[nevents]
        
enrolled = ifelse(enrollment_times < threshold_power[k,2], 1, 0)
        
survival_time = ifelse(pseudo_surv > threshold_censoring[k,2],
threshold_censoring[k,2] - enrollment_times, event_time)
        
censor = ifelse(pseudo_surv > threshold_censoring[k,2], 0, 1)

data.df2 <- data.frame(survival_time,censor,group,enrolled)
        
data.df2 = data.df2[data.df2[,4]==1,]
        
        
fit.wlr00_st2 <- wtdlogrank(Surv(survival_time, censor)~group, 
data = data.df2, sided = 1, WtFun="FH", param=c(prho,pgamma))
        
reject[k,2] = ifelse(abs(fit.wlr00_st2$Z) > rejectionBoundary2, 1, 0)


\end{lstlisting}

\end{document}